\documentclass[twoside,english]{article}
\usepackage[super,sort&compress,comma]{natbib}

\usepackage[version=3]{mhchem}

\usepackage[left=2cm, right=2cm, top=2cm, bottom=2.0cm]{geometry}
\usepackage{graphicx} 

\usepackage[small]{caption}
\usepackage[english]{babel}
\usepackage[T1]{fontenc}
\usepackage[utf8]{inputenc} 
\usepackage[colorlinks=true, urlcolor=blue, linkcolor=blue, citecolor=blue]{hyperref}

\usepackage{doi}

\newcommand{\bref}[1]{(\ref{#1})}
\usepackage{siunitx}
\usepackage{makecell}

\usepackage[normalem]{ulem} 
\usepackage{xcolor}

\usepackage{tikz}
\usetikzlibrary{arrows}
\usetikzlibrary{decorations.markings}
\usetikzlibrary{fadings,shapes.arrows,shadows}   

\newcommand*{\defeq}{\mathrel{\vcenter{\baselineskip0.5ex \lineskiplimit0pt
                     \hbox{\scriptsize.}\hbox{\scriptsize.}}}%
                     =}

\newcommand{\braces}[1]{\left(#1 \right)}
\newcommand{\cbraces}[1]{\left\{#1 \right\}}
\newcommand{\vbraces}[1]{\left|#1 \right|}
\newcommand{\gbraces}[3]{\left #1 #2 \right #3}

\renewcommand{\vec}[1]{\boldsymbol{#1}}
\newcommand{\myvec}[1]{\boldsymbol{#1}}

\begin{document}

\title{
Quantitative modelling of nutrient-limited growth of bacterial colonies in microfluidic cultivation
}
\maketitle 
\noindent
Raphael Hornung,\textit{$^{a\ddag}$} 
Alexander Gr\"unberger,\textit{$^{bc\ddag}$}
Christoph Westerwalbesloh,\textit{$^{b}$}
Dietrich Kohlheyer,\textit{$^{bd}$}
Gerhard Gompper,\textit{$^{a}$}
Jens Elgeti\textit{$^{*a}$}

\noindent
\textit{$^{a}$Theoretical Soft Matter and Biophysics, Institute of Complex Systems and Institute for Advanced Simulation, Forschungszentrum J\"ulich and JARA, 52425 J\"ulich, Germany. }

\noindent
\textit{$^{b}$Institute of Bio- and Geosciences, IBG-1: Biotechnology, Forschungszentrum J\"ulich, 52425, J\"ulich, Germany.}

\noindent
\textit{$^{c}$Multiscale Bioengineering, Bielefeld University, Universit\"atsstr. 25, Bielefeld, 33615, Germany.}

\noindent
\textit{$^{d}$Aachener Verfahrenstechnik (AVT.MSB), RWTH Aachen University, 52056, Aachen, Germany.}

\noindent
\textit{$^{\ddag}$These authors contributed equally to this work.}

\noindent
\textit{$^{*}$Corresponding author, E-mail: j.elgeti@fz-juelich.de}

\section*{Abstract}
Nutrient gradients and limitations play a pivotal role in the life of all microbes, both in their natural habitat as well as in artificial, microfluidic systems. Spatial concentration gradients of nutrients in densely packed cell configurations may locally affect the bacterial growth leading to heterogeneous micropopulations. A detailed understanding and quantitative modelling   of cellular behaviour under nutrient limitations is thus highly desirable. We use microfluidic cultivations to investigate growth and microbial behaviour of the model organism \textit{Corynebacterium glutamicum} under well-controlled conditions. With a reaction-diffusion type model, parameters are extracted from steady-state experiments with a one-dimensional nutrient gradient. Subsequentially, we employ particle-based simulations with these parameters to predict the dynamical growth of a colony in two dimensions. Comparing the results of those simulations with microfluidic experiments yields excellent agreement. Our modelling   approach lays the foundation for a better understanding of dynamic microbial growth processes, both in nature and in applied biotechnology.

\noindent
\newline
\textbf{Keywords}: nutrient gradients $\cdot$ bacterial nutrient consumption $\cdot$ growth simulations $\cdot$ microfluidic cultivation 

\twocolumn

\section{Introduction}Growth of cells is enabled by diffusive factors, a property which is universal for all living 
processes ranging from growth of single bacteria \cite{Wang2010,Taheri-Araghi2015} or eucaryotic 
cells \cite{gospodarowicz1976growth,Hirschhaeuser2010} to tissue \cite{Rumpler1173,Lovett2009,Huh2011,Bhatia2014} 
and biofilm formation. \cite{donlan2002biofilms, stewart2003diffusion,Wilking2011,Liu2015} There is a complex interplay between diffusion and uptake, strongly influenced by metabolism and environment. Typically, various environmental perturbations -- 
such as nutrient gradients, oxygen depletion, temperature changes, and others -- occur simultaneously, 
rendering the analysis of nutrient limitations and their influence on individual cellular systems challenging.
Microfluidic cultivation systems - which often consist of microfluidic growth chambers with well-controlled nutrient supply  - are ideal to analyse and quantify cellular behaviour under defined environmental conditions. \cite{whitesides2006origins, grunberger2014single} They facilitate the investigation of single selected limiting factors while keeping others 
in a defined range. \cite{schmid2010chemical,kortmann2010single} Furthermore, microfluidic cultivation 
enables observation of cellular growth patterns by microscopy with a high spatio-temporal resolution. 
This combination allows quantitative modelling   and parameter extraction. Whereas most microfluidic studies 
so far have used undefined or defined but non-limiting conditions for the cultivation, only few studies 
have artificially limited cell growth within microfluidic devices.
\cite{mather2010streaming,C5LC00646E,Cherifi2017Restriction}
\begin{figure*}[t]
\begin{center}
\includegraphics[width=\textwidth]{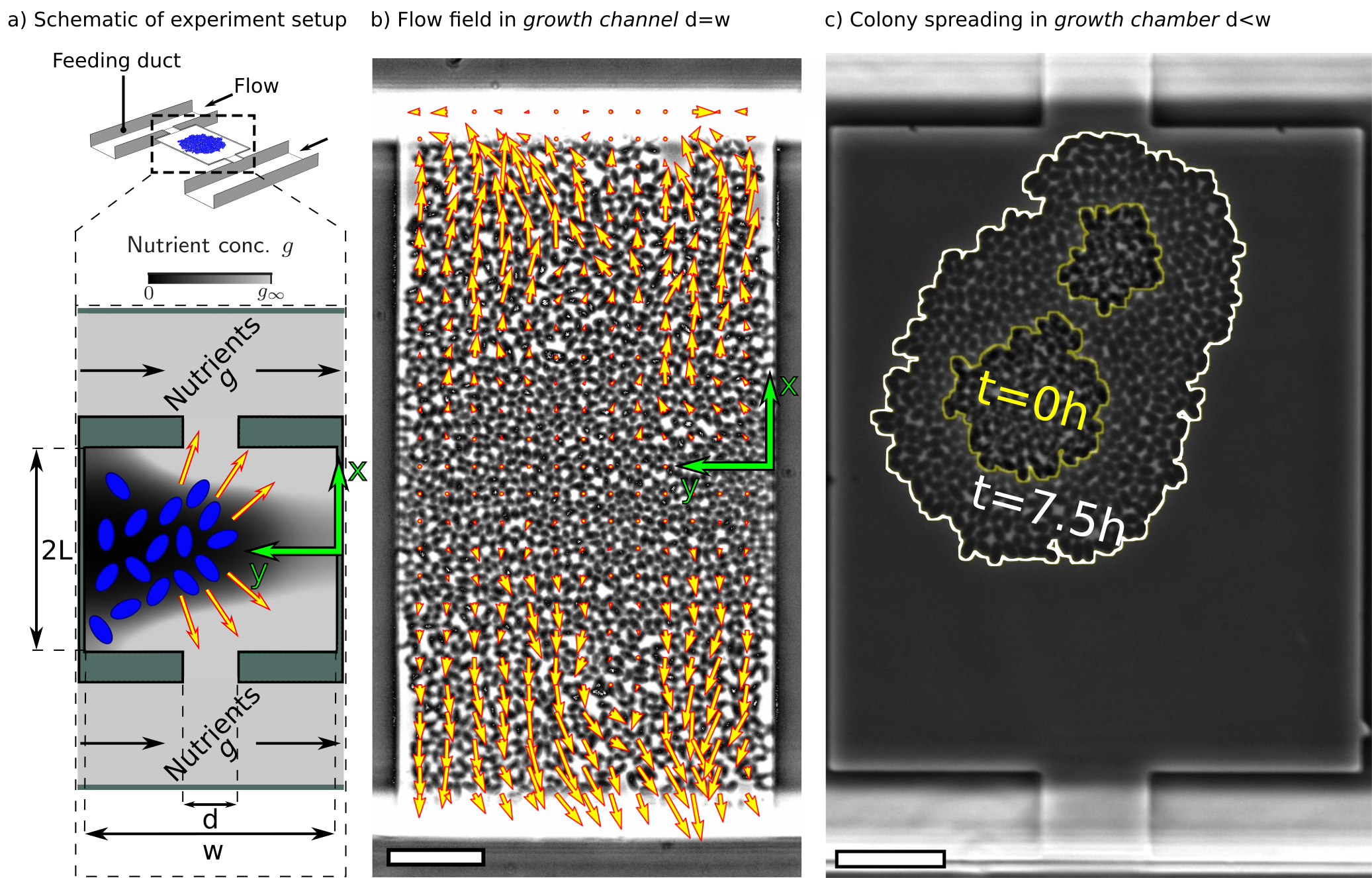}
\end{center}
\caption{{\bf Microfluidic setup.} {\bf (a)} Schematic of microfluidic experiment in which \textit{Corynebacterium glutamicum} are grown. The microfluidic device provides constant nutrient supply by the flow in the large feeding ducts; $g_{\infty}$ denotes the feeding concentration. Inside the chamber the bacteria (blue) take up the nutrients, the local concentration $g$ drops and nutrient gradients occur (greyscale depicting local concentration). The bacteria grow and a flow towards the channel outlets evolves (red-yellow arrows). {\bf (b)} Snapshot of the \textit{growth channel} setup used to observe the steady-state flowfield of bacteria. Overlayed vectors depict the flowfield of bacteria inside the channel, as measured by Particle Image Velocimetry (PIV). Scale bar  length \SI{10}{\micro\meter}.
{\bf (c)} Overlay of two snapshots at different timepoints of an experiment in a \textit{growth chamber} with narrow outlets to study the spreading dynamics of a bacterial colony. The yellow and white lines depict the perimeter of the colony at different timepoints. Scale bar length \SI{10}{\micro\meter}.}
\label{fig:setup}
\end{figure*}
To understand how nutrient limitation affects growth patterns, we investigate the growth of the bacterium 
\textit{Corynebacterium glutamicum}, a non-motile bacterial model organism, at different distinct carbon-source 
concentrations. Our microfluidic system \cite{grunberger_disposable_2012} acts as a ``quasi'' chemostat and allows the 
investigation of growing microcolonies in a two-dimensional monolayer under well-defined 
environmental conditions (Fig.~\ref{fig:setup}). 
In a first step, we grow cells in a quasi-one-dimensional setup, i.e. a wide and open \textit{growth channel} (Fig.~\ref{fig:setup}(b)). We derive a minimal theoretical model for growth which we fit to the steady-state motion pattern of cells for a single experimental condition. Our colony-growth model allows a direct read-off of the 
nutrient-uptake function from experimental data; the results show good agreement with 
Monod \cite{Monod1949} and Teissier \cite{teissier1937lois} nutrient-uptake functions.
With the fit we extract the length scale of nutrient depletion within a colony and the nutrient dependency of growth. 
In a second step, we extrapolate to other feeding concentrations. 
Encouraged by excellent agreement, in a third step we extrapolate from this steady-state, quasi-one-dimensional 
geometry to the prediction of time-dependent growth of bacteria in a genuine two-dimensional setup (\textit{growth chamber}, Fig.~\ref{fig:setup}(c)). To do so, we feed the fitted parameters into a particle-based growth simulation 
\cite{basan_dissipative_2011-1} and compare the shape and area of growing colonies over time. 
The simulations show striking agreement with growth dynamics observed experimentally, demonstrating 
the predictive power of our minimal modelling   approach.
Currently, measurements of concentration profiles on the microscale are unfeasable. Our findings close this gap by providing a way to quantify nutrient distributions from the measurable velocity profile. Furthermore, our results serve as a basis for the design of optimised microfluidic systems for microbial cultivation.
\section{Results and Discussion}
\subsection{Theory for Nutrient Uptake and \\ Biomass Conversion}
Cells need to metabolise nutrients in order to grow. The amount of nutrients taken up depends on the concentration $g=\hat{g} \bar{g}$ of nutrients available, where we introduced the dimensionless concentration $\hat{g}$ and the unit conversion factor $\bar{g}$. Furthermore, it is reasonable to assume that the nutrient uptake rate $u$ per bacterium
has an upper bound $u_\infty\bar{g}$. Hence, we write the total nutrient uptake rate per bacterium as  
$u_\infty\bar{g} u\!\braces{\hat{g}}$, where $u\!\braces{\hat{g}}$ is a dimensionless function varying between zero and unity. The nutrient concentration field $\hat{g}$ thus obeys
\begin{equation}
\label{eq::base_model1}
\partial_t \hat{g} = D \Delta \hat{g} - \varrho u_\infty u\!\braces{\hat{g}}, 
\end{equation}
where we assume the uptake to be linear in bacterial number density $\varrho$ and only diffusive nutrient transport \cite{C5LC00646E} with diffusion constant $D$. 
Bacteria convert the nutrients absorbed into biomass with efficiency $\epsilon$, i.e. cells grow with a rate of 
$k=\epsilon u_\infty u\!\braces{\hat{g}}$. 
Thus the bacterial density evolves according to
\begin{equation}
\partial_t \varrho = -\nabla\cdot\braces{\varrho \vec{v}} + \varrho \epsilon u_\infty u\!\braces{\hat{g}} \, ,
\label{eq::base_model2}
\end{equation}
with the bacteria flow velocity $\vec{v}$. The efficiency parameter $\epsilon$ describes essentially the amount of nutrients a bacterium needs to consume in order to divide. Similar reaction-diffusion type models have also been used for example to investigate velocity and shape of bacterial growth fronts \cite{Farrell2013} or the growth behaviour of bacterial aggregates.\cite{melaugh2016shaping} For constant bacterial density and fast diffusion these equations simplify to
\begin{eqnarray}
\Delta \hat{g} =  u\!\braces{\hat{g}}/l_g^2  \ \ , \ \ 
&\text{ }&\nabla\cdot\vec{v}= \epsilon u_\infty u\!\braces{\hat{g}} \, ,
\label{eq::base_model2_ss}
\end{eqnarray}
with two parameters, the nutrient decay length $l_g = \sqrt{ D/\varrho u_\infty}$ describing the ratio of nutrient-uptake rate and diffusive nutrient flux, and the maximum growth rate $k_{max} = \epsilon u_\infty$. Furthermore, the shape of the 
uptake function $u\!\braces{\hat{g}}$ determines at which nutrient-concentration-scale limitation of growth occurs. 
The exact uptake rate as a function of nutrient availability $u\!\braces{\hat{g}}$ is generally not known; however, 
in one dimension, model eqns.~\bref{eq::base_model2_ss} can be rearranged to read off $u$ from a given velocity profile $v\!\braces{x}$. For a quasi-one-dimensional colony of length $2L$, with prescribed nutrient concentration $\hat{g}(\pm L)=\hat{g}_\infty$ at the boundaries, elimination of $u\!\braces{\hat{g}}$ from eqns.~\bref{eq::base_model2_ss} yields
\begin{equation}
\hat{g}\!\braces{x} = \braces{V(x)-V_0}/\braces{k_{max}l_g^2} ,
\label{eq::g_V}
\end{equation}
where $V\braces{x} = \int_{x_0}^x v\braces{x'}dx'$ the integral of $v$ starting from the flow symmetry axis $x=x_0$ (at which $g'\braces{x_0}=v\braces{x_0}=0$) and $V_0 = k_{max}l_g^2\hat{g}\braces{x_0}$. 
Insertion of eq.~\bref{eq::g_V} into eq.~\bref{eq::base_model2_ss} eliminates the nutrient concentration, so that
\begin{equation}
\frac{dv}{dx} = k_{max} u\!\braces{ \braces{V(x)-V_0}/\braces{k_{max}l_g^2}  } \, . 
\label{eq::dvdx_V}
\end{equation}
Thus the shape of the uptake function $u(\hat{g})$, and the scalar parameters $l_g$ and $k_{max}$ can be extracted from a fit of our model to experimental results for a quasi-one-dimensional system, as will be discussed below.
\subsection{Growth Channel Experiments with \textit{Corynebacterium glutamicum}}
\begin{figure}[!t]
\begin{center}
\includegraphics[width=\linewidth]{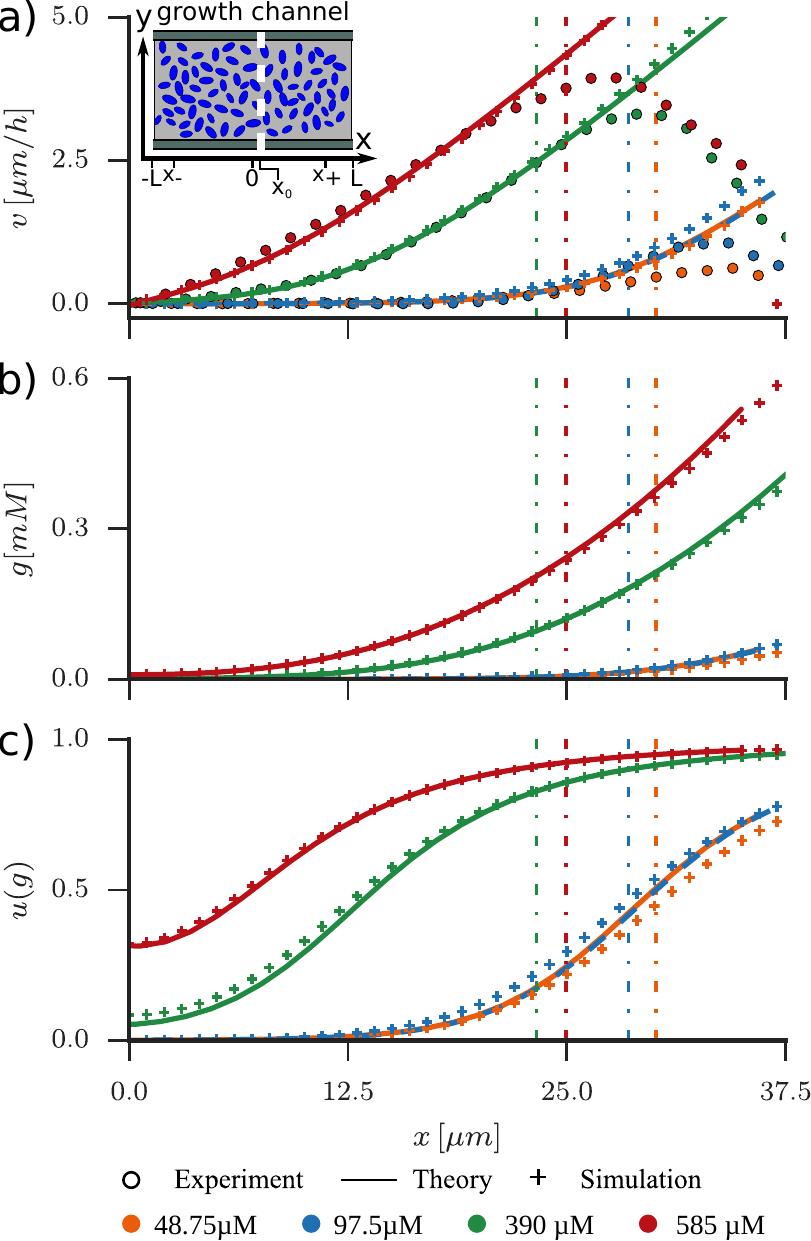}
\caption{{\bf Velocity and nutrient concentration profile in growth channel setup.} 
{\bf (a)} Velocity profiles estimated by PIV of growth channel experiments and subsequent averaging over
the $y$-direction and time (circles). One example for
each feeding concentration $g_{\infty}$ is shown, as indicated in legend (see Fig.~S1 for all data). Due to the symmetry of the velocity profile with respect to the channel center, the $x$-range from the channel center to the feeding outlets is displayed. The flow symmetry axis position $x_0$ has also been fitted to account for small deviations from the channel center at $x=0$. Dashed-dotted vertical lines indicate the cutoff $x_+$ used to
constrain the data range used for the fit to our analytic model
eqns.~\bref{eq::base_model2_ss} (for details of fit see \hyperref[ss:fitProc]{Materials and Methods}). 
Continuous lines show the velocity profiles of the model fit using Monod-uptake which has been extrapolated towards the channel outlets. The line for the concentration $g_{\infty} = 48.75$ is dashed to increase visibility. Flow profiles of corresponding particle-based simulations are shown with ``+''-symbols. Experimental and simulation data has been averaged over time and $y$-direction.
{\bf (b)-(c)} $g$- and $u\!\braces{g}$-profiles from model fit (continuous lines) and corresponding particle simulation results (``+''-symbols).
}
\label{fig::oneD_flowprof_gconc}
\end{center}
\end{figure}
We investigated the growth of \textit{C. glutamicum} in a quasi one-dimensional \textit{growth channel} geometry (see Fig.~\ref{fig:setup}(a)-(b)) to facilitate modelling  and interpretation of growth patterns. To prevent co-metabolism of different carbon sources \cite{unthan_beyond_2014}, modified CGXII medium without glucose was used as growth medium. Here protocatechuic acid (PCA) served as sole carbon source and growth limiting factor. 
We performed growth channel experiments with four different concentrations in the feeding duct (see Fig.~\ref{fig:setup}(a)) $g_{\infty}=\SI{48.75}{\micro M} \braces{n=2\text{ independent experiments}},\SI{97.5}{\micro M}$ $\braces{n=3}$, $\SI{390}{\micro M} \braces{n=3}$ and $ \SI{585}{\micro M} \braces{n=4}$ PCA in aqueous solution. Time-lapse phase contrast microscopy images of the growing microcolonies were recorded every $\Delta t = 5-10\si{\minute}$ over 40h of microfluidic cultivation to follow the growth on different feeding levels. Starting from a few bacteria seeded into the growth channel, bacteria grow, divide and populate all available space in the growth channel. Bacteria are pushed out of the channel into the feeding duct and are dragged away by the flow. Finally a continuous steady state evolves which is characterised by the balanced growth and outflow of bacteria (see movie M1).
Nutrient limitation was clearly visible for the lowest feeding concentrations $\SI{48.75}{\micro M}\text{ and }\SI{97.5}{\micro M}$ where we observed an almost complete growth arrest in $\approx 2/3$ of the growth channel (see movie M2).
This decline of growth activity clearly demonstrates the presence of nutrient gradients which develop on a length scale of a few cell sizes. At the same time, for the highest feeding concentration $\SI{585}{\micro M} \braces{n=2}$, no growth arrest zones were visible and bacterial biomass production took place along the whole channel length, paralleled by a strong flow of cells towards the channel outlets. 
For the intermediate feeding concentration of $\SI{390}{\micro M}$ we observed a mixed picture: the flow was clearly much stronger than for the lowest feeding concentration, but a small fraction close to the chamber center exhibited low to no growth. 
In order to quantify the observed growth patterns, we analysed the flow patterns $\vec{v}=\braces{v_x,v_y}$ using 
Particle Imaging Velocimetry \cite{raffel2013particle} (PIV, see \hyperref[matmet::imaging_and_piv]{Materials and Methods}). 
In steady state, we observed a plug-like flow with almost no dependence of the velocity on the $y$-position. 
To compare the velocity profile with our theory, we define $v\braces{x}$ as the average of $v_x$ along the $y$-direction and over all steady-state timepoints. 
Resulting velocity profiles $v$ for the different feeding levels are shown in Fig.~\ref{fig::oneD_flowprof_gconc}(a).
The growth arrest zones are clearly visible.
At the channel outlets, PIV underestimates the velocity due to (I)
bacteria being washed out of the channel such that the correlation
with the next frame often fails and (II) because velocity and
frequency of division events increase, both of which raise the
difficulty of a correct correlation match. 
Thus, the velocity decrease close to the channel outlets and the
corresponding maxima are artefacts. We decided to limit our quantitative
analysis to a central region of the channel heuristically defined as
the interval between the two inflection points closest to the two maxima
at the channel outlets ($x_\pm$, see Fig.~\ref{fig::oneD_flowprof_gconc}(a)-(c),
vertical dashed-dotted lines). We also checked the sensitivity of our results to include all data points up to the velocity maxima near the channel outlets and found only minor deviations.
\subsection{Matching Continuum Model and Experiments}
\begin{figure}[!t]
\centering
\includegraphics[width=\columnwidth]{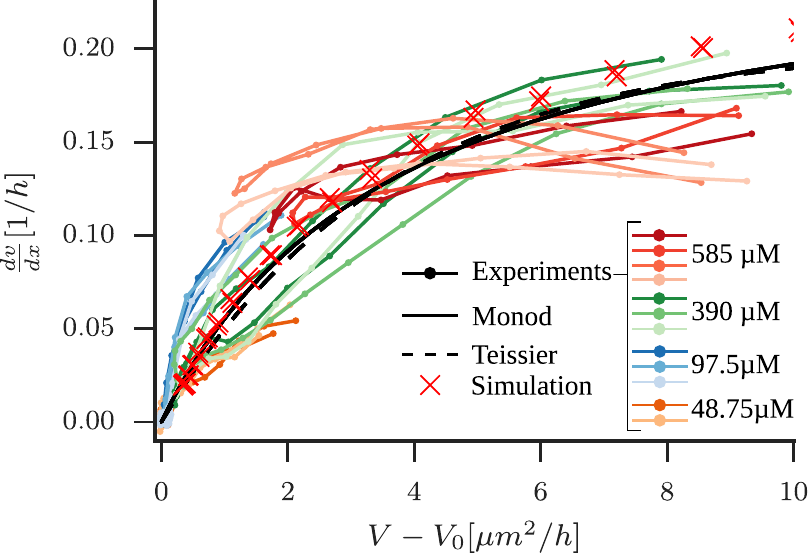}
\caption{{\bf Uptake function extracted from PIV data}. A plot of  $\frac{dv}{dx}$ versus $V=\int_{x_0}^x v\braces{x'}dx'$, the integral of $v$ starting from the flow symmetry axis at $x=x_0$, for different nutrient concentrations $g_\infty$, as indicated in legend. Curves of the different experiments $i$ have been shifted by $V_0^i = k_{max}l_g^2\hat{g}^i\braces{x_0}$ according to eq.~\bref{eq::dvdx_V}, where the $x_0^i$ and $V_0^i$ have been estimated by a least-square fit (details of fit described in \hyperref[ss:fitProc]{Materials and Methods}).  
Black lines show fits with Monod-uptake $u=\hat{g}/\braces{1+\hat{g}}$ (continuous) and Teissier-uptake $u =1-\exp\braces{-\hat{g}}$ (dashed), the red ``x'' result of a corresponding particle-based simulation. Experimental and simulation data has been averaged over time and the $y$-direction.}
\label{fig:u}
\end{figure}

We commence our analysis with the estimation of a suitable uptake function $u$ from our velocity data. 
Relation~\bref{eq::dvdx_V} predicts that a $dv/dx-V$-plot of our measured velocity profiles collapses onto 
the uptake function $u$ if each curve is shifted along the $V$-axis by an offset $V_0=k_{max}l_g^2 \hat{g}\braces{x_0}$ 
proportional to the concentration at the starting point $x_0$ of the integration (see Fig.~\ref{fig:u}). 
An initial guess for the shifts $V_0^i$ is obtained easily by visual inspection, since continuity demands that 
data sets for $dv/dx-V$ from different experiments have to overlap.
Different models for uptake kinetics \cite{Monod1949,teissier1937lois} agree on a set of conditions:
\begin{flalign}
\label{eq::u_lin}
& \bullet\text{linear for small concentrations } u\!\braces{\hat{g}} \overset{\hat{g}\rightarrow 0}{\propto} \hat{g},\\
\label{eq::u_sat}
& \bullet\text{saturating at high concentrations } u\!\braces{\hat{g}} \xrightarrow{\hat{g}\rightarrow \infty} 1 ,\\
\label{eq::u_mon}
& \bullet\text{monotonically increasing } u' > 0 \text{ and}\\
\label{eq::u_curv}
& \bullet\text{concave everywhere, i.e. } u'' < 0.
\end{flalign}
which are also consistent with our data.
We choose the unit conversion factor $\bar{g}$ such that the linear slope for small concentrations in condition \bref{eq::u_lin} is equal to unity. Thus, the length scale $l_g$ describes the exponential decay length of the nutrient concentration under limiting conditions, where $u\braces{\hat{g}}\approx\hat{g}$. The conditions~\bref{eq::u_lin} and \bref{eq::u_sat} imply a simple geometrical interpretation of the parameters $k_{max}$ and $l_g$ in the $dv/dx-V$-plane. From eq.~\bref{eq::dvdx_V} and condition~\bref{eq::u_lin} it follows $dv/dx \overset{\hat{g}\rightarrow 0}{=} \braces{V-V_0^i}/l_g^2$, eq.~\bref{eq::dvdx_V} and condition~\bref{eq::u_sat} yield $dv/dx \overset{\hat{g}\rightarrow\infty}{=} k_{max}$. Hence, $k_{max}$ and $l_g$ fix the initial slope and the saturation value of the $dv/dx-V$-curve. 
We use two common uptake models which comply with conditions \bref{eq::u_lin}-\bref{eq::u_curv}, Monod-uptake $u\!\braces{\hat{g}}=\hat{g}/\braces{1+\hat{g}}$ \cite{Monod1949} is explored here (see Fig.~\ref{fig:u}), the very similar results for Teissier-uptake $u\!\braces{\hat{g}}=1-\exp\braces{-\hat{g}}$ \cite{teissier1937lois} can be found in Fig.~S2.
With a suitable uptake function $u\!\braces{\hat{g}}$ our model, defined by eqns.~\bref{eq::base_model2_ss}, 
is complete and the theoretical predictions can be fitted to the experimental data.
A direct fit of $u$ in the $dv/dx - V$ diagram, using eq.~\bref{eq::dvdx_V} to estimate the parameters $k_{max}$ and $l_g$, 
provides an initial estimate. However, the numbers are error-prone due to the derivative of noisy experimental data. 
Thus, we fit the solution of eqns.~\bref{eq::base_model2_ss} for the velocity profile $v\braces{x}$ 
to the measured flow profiles with $k_{max}$ and $l_g$ as fit parameters. 
To account for small deviations of the velocity symmetry axis position from the channel center, we employ the concentration at the center and the center position as additional fit parameters (see \hyperref[ss:fitProc]{Materials and Methods} for details).
Note that in Fig.~\ref{fig:u} the experiments at $g_{\infty} = \SI{390}{\micro M}$ span almost the full relevant concentration range; thus, flow profiles $v\braces{x}$ are fitted for this concentration and the model is then used to predict the velocity profiles at higher and lower concentrations.
Model fits give good predictions of experimental data, even for extrapolations to very different concentrations 
(see Fig.~\ref{fig::oneD_flowprof_gconc}(a)). The corresponding fitted Monod and Teissier uptake functions are depicted in 
Fig.~\ref{fig:u}. For the positions $x$ outside the interval $x_- \le x \le x_+$ used for the fit (see Figs.~\ref{fig::oneD_flowprof_gconc} and S1, continuous model curves outside dashed-dotted lines), model predictions agree reasonably well up to the velocity maxima. 
Our model also predicts the nutrient concentration profile $\hat{g}(x)$ inside the channel (see Fig. \ref{fig::oneD_flowprof_gconc}(b)-(c)); however, we emphasise that our fit procedure does not prescribe the feeding concentrations $g_\infty$ as present in the experiments. Comparision of the feeding concentrations via the linear relation $\hat{g}_\infty = g_\infty/\bar{g}$ thus provides an additional consistency check. Linear fits (see Fig.~\ref{fig:gbarEst}) match well and estimate the conversion factor to $\bar{g}=0.02 \si{\milli M}$ for both Monod and Teissier-uptake.
Note that the theoretical feeding concentration $\hat{g}_\infty$ is estimated from an extrapolation of the $\hat{g}$-profile 
to the channel outlets (for details see \hyperref[ss:fitProc]{Materials and Methods}). Due to the reduced bacterial density, our model is not strictly valid at the channel outlets; thus, this approach only provides an estimate of the concentration scale $\bar{g}$.
We define the concentration scale $g_{1/2}$ as the nutrient concentration at which uptake and growth rates are half of their maximum values, thus for $\hat{g}_{1/2}$ holds $u\!\braces{\hat{g}_{1/2}}=1/2$. With the concentration scale $\bar{g}$ we estimate $g_{1/2}$ for Monod and Teissier uptake around $g_{1/2}\approx 13-\SI{20}{\micro M}$, which is about five to ten percent of the PCA-concentration in standard CGXII medium; much lower than we previously assumed for Monod kinetics ($g_{1/2}=\SI{100}{\micro M}$).\cite{unthan_beyond_2014,C5LC00646E} 
Our observed maximal growth rate of around $k_{max}\approx 0.2-\SI{0.26}{\hour^{-1}}$ agrees well with previous observations.\cite{unthan_beyond_2014} The nutrient decay length of $l_g\approx 3.8-\SI{4.2}{\micro\meter}$ falls in the range of $l_g\approx\SI{2}{\micro\meter}-\SI{5}{\micro\meter}$ which can be estimated from previous results,\cite{unthan_beyond_2014} which are affected, however, by large uncertainties. In particular, the conversion of uptake rates measured per gram cell dry weight $\braces{g_{CDW}}$ into uptake per single cell is prone to large errors, since reported single-cell weights vary by an order of magnitude.\cite{C5LC00646E}  
However, it is important to emphasise that the parameter estimates of previous studies\cite{unthan_beyond_2014} were based on the assumption of spatial homogeneity. Our main result here is that nutrient gradients are very important and have to be considered. Thus, it is no surprise that estimates differ, and we deem the approach of the present study to be more reliable (see Table \ref{tab:fitResP} for all fit results).
\subsection{Spreading Dynamics of \textit{C. glutamicum}}
\label{subsection::theory_vs_exp}
\begin{figure*}[!t]
\begin{center}
\includegraphics[width=5.5in]{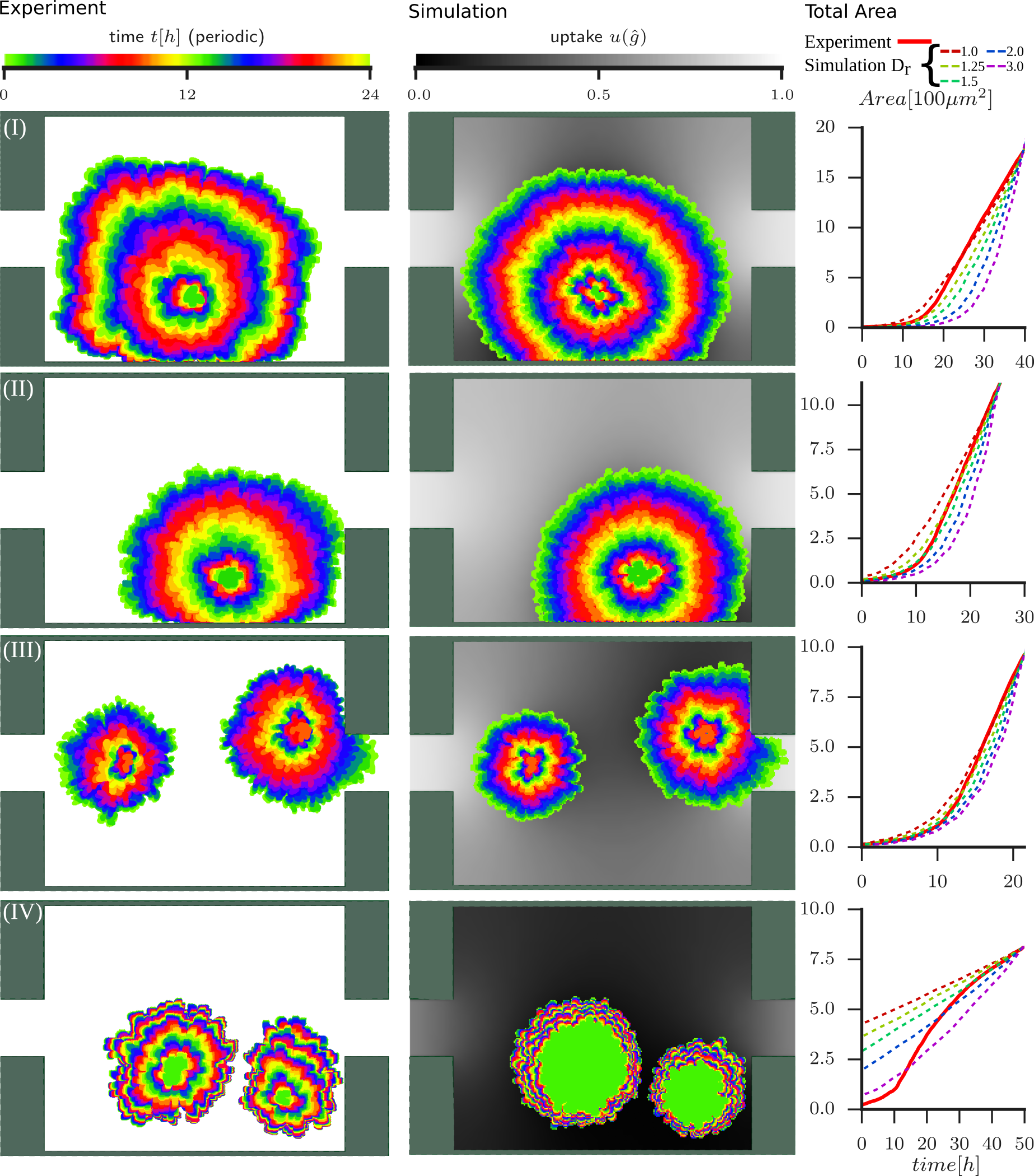}
\caption{{\bf Dynamics of 2D colony growth.} Four examples of colony spreading dynamics in simulation and experiment (see also movies M3-M6). Rows \textbf{(I)-(III)} belong to experiments with $g_\infty = \SI{195}{\micro M}$ while \textbf{(IV)} shows an experiment with feeding concentration of $g_\infty = \SI{19.5}{\micro M}$. \textbf{Left column:} Depiction of the colony shape dynamics in experiments. The outlines of the colony at equidistant timepoints are shown in different colours with a periodic colourscale (legend on top), 12 hours have passed between two rings with the same colour. \textbf{Center column:} Same depiction as in left column for outlines from the corresponding simulation with $D_{free}/D_{bulk}=1.25$ (colourscale for outlines of colony is the same as in left column). The greyshaded area around the colonies shows the profile of $u\!\braces{g}$ at the last timepoint, illustrating the local limitation of growth due to nutrient depletion (legend on top). \textbf{Right column} shows comparison of total colony area $A\braces{t}$ in experiment (red line) and five corresponding simulations (dashed lines) with $D_r \defeq D_{free}/D_{bulk}=1,1.25,1.5,2,3$. The simulations have been shifted along the time axis such that they cross the last data point of the experiment (see Fig.~S5).}
\label{fig::ov_exp_sim}
\end{center}
\end{figure*}
We adapted a particle-based simulation technique for growing tissues \cite{basan_dissipative_2011-1} to quantitatively predict
bacterial growth and nutrient distribution inside arbitrarily shaped growth chambers.
In short, each bacterium is represented by two point particles that repel each other by a growth force. After a critical size 
threshold is reached, the cell divides, and two new particles are added. To model the feedback on growth, the growth force is taken to be proportional to the nutrient uptake. Nutrients are supplied with constant density at the outlets and are consumed by the bacteria (for details of simulation technique see \hyperref[matmet::simulation]{Materials and Methods}). 
This model and simulation technique is also able to incorporate features currently not taken into account, such as the 
finite elasticity of bacteria or pressure-dependent growth, offering many possibilities for further studies. 
To demonstrate the predictive power of our simulation model, we calculate the time-dependent growth of bacteria inside a 
rectangular \textit{growth chamber} with two narrow feeding outlets at both sides, as depicted in Fig.~\ref{fig:setup}(a) and (c), and compare our predictions with experimental results. 
In the experiments, we analyse spreading of a colony of \textit{C. glutamicum} with prescribed 
feeding concentrations $g_\infty = \SI{19.5}{\micro M}\,(n=3)$ and $g_\infty = \SI{195}{\micro M}\,(n=7)$. 
A few bacteria are seeded in the growth chamber at $t=0$, and subsequent colony spreading is observed via time-lapse imaging. 
Clearly, the spreading dynamics is very sensitive to the initial conditions. If initially all bacteria are concentrated
at one spot, a single circular colony develops, whereas if the bacteria are initially distributed over the chamber, 
multiple separate colonies grow and finally merge (see left column in Figs.~\ref{fig::ov_exp_sim} and S3). The simulations are initialised with the same amount of cells at identical positions as in the experiment.
We convert the experimental feeding concentration $g_\infty$ to simulation units via the previously calculated concentration scale $\bar{g}$. Visual comparison of the shape of the colony predicted by simulations and observed in experiments already shows excellent agreement, with growth patterns in nice synchrony (see center column in Fig.~\ref{fig::ov_exp_sim}), especially for colonies larger than  $A=\SI{500}{\micro \meter^2}$. 
Furthermore, the overall colony area $A\braces{t}$ over time serves as an easily accessible quantifier for comparison
(see Fig.~\ref{fig::ov_exp_sim}, right column). The growth of colonies consisting of only a few bacteria depends strongly on the state of the cell cycle of every individuum, such that we expect a large variability in growth. Therefore, we expect that our model agrees best in the later stages of the experiment, when memory effects have worn out 
and the continuum description is appropriate. The simulation time axis is therefore shifted such that $A(t)$ coincides with the last data point of the experiment.
For experiments with concentration $g_\infty = \SI{195}{\micro M}$ (i.e. half our ``fitting concentration''), 
simulations agree very well with experiments even down to colonies consisting of only the few cells at the starting 
point of the experiment (see Fig.~\ref{fig::ov_exp_sim}(I)-(III)).  
When extrapolating to the much lower concentration $g_\infty = \SI{19.5}{\micro M}$ 
(see Fig.~\ref{fig::ov_exp_sim}(IV)) simulations still agree remarkably well for colonies larger than about 
$\SI{500}{\micro \meter^2}$. In experiments with very low nutrient concentrations ($g_\infty$ = \SI{19.5}{\micro M}), bacteria initially grow faster than predicted. This may be due to cell-history effects from preculture, e.g. carbon storage, or effects from the differences in population densities present in growth-channel experiments versus the smaller initial density in growth-chamber experiments. A quantitative understanding of this effect requires a more detailed study in the future.

Furthermore, our observations suggest directed growth toward the channel inlets. 
While, initially, colonies grow in a more or less circular shape, some elongate over time. 
The effect is weak, and sometimes caused by previous wall contact. However, in some cases (right colony in Fig. 4
(IV) and Fig. S3 (II) and (IV)) orientation toward the inlets is visible. This indicates that the nutrient gradients, as predicted by the local growth profile u(g) (see Fig. 4, center column) of our model, directly affect the temporal expansion of the
colony. However, these changes in shape are subtle and further quantitative analysis is required.

Simulations can be improved further when hindered diffusion through bacteria is considered. In our model, 
this can be accounted for in a coarse-grained manner by defining two different diffusion constants, $D_{bulk}$ and
$D_{free}$ in- and outside of the colony, respectively. For spatially changing diffusion coefficient, the $D\Delta g$ term in eq.~\bref{eq::base_model1}, has then to be replaced by $\nabla\cdot\braces{D\nabla g}=D\Delta g + \nabla D\cdot \nabla g$. The agreement with experiments is best for $D_{free}/D_{bulk}\approx 1.25$, a surprisingly small impedement if the bacteria were considered as obstacles.\cite{ochoa1994diffusive,wood2002calculation}
\section{Conclusion}
In summary, we have shown how microfluidic devices can be exploited to measure the effect of 
nutrient availability and limitation on growth and transport in bacterial microcolonies. 
With some simple assumptions, like diffusive transport and mass balancing, we are able to 
quantitatively model growth and uptake kinetics.
Our results show that at low nutrient concentration gradients in growth develop rapidly
after a critical cell-colony size is reached (Fig.~\ref{fig::ov_exp_sim}), both in experiment 
and simulation. It would be interesting to extend such studies to growth dynamics in the presence of antibiotics. \cite{allen2016antires} Furthermore, this approach should also work for eucaryotic and even mammalian cells and cell-lines, where nutrient limitation can be of pivotal importance.\cite{Hirschhaeuser2010}
The modelling framework presented here can also be used and extended to optimise microfluidic 
geometries to guarantee and maintain optimal nutrient supply.\cite{C5LC00646E} In particular, 
we hope that our results will stimulate a discussion about the existence and influence of 
environmental gradients within microfluidic cultivation systems.
Our results can also serve as a basis for studies in related fields, such as the investigation 
of mechanical forces within cell growth and development.\cite{montel_stress_2011,Grant20140400,Podewitz2015,Delarue2016,Farrell2017}
Here, our approach could be used to quantitatively determine the effect of nutrient transport and 
limitation in order to extract the contribution of mechanical forces.
\section{Materials and Methods}
\subsection{Microbial Strain and Cultivation}
\label{matmet::strain_and_medium}
Model organism in this study was \textit{C. glutamicum} wild type (ATCC 13032). Microfluidic cultivations were performed using two different cultivation chamber systems. For steady state growth a monolayer growth chamber as described by \cite{mather2010streaming} was used, with a chamber dimension of $40\times 75 \times 1 \si{\micro\meter}$. For dynamical growth studies, the cultivation system described by \cite{grunberger2015spatiotemporal} was applied. For chip fabrication, details and further information the reader is referred to \cite{gruenberger2013microfluidic,grunberger2015spatiotemporal}. To prevent co-metabolism of different carbon sources \cite{unthan_beyond_2014}, modified CGXII medium without glucose was used as growth medium. Here protocatechuic acid (PCA) serves as a sole carbon source and was varied in concentration within the different sets of experiments. Modified CGXII medium was infused at approx. $\num{200}\,\si{\nano\litre\per\minute}$ after cell inoculation. Phase equilibrium experiments where performed as followed. First cells were cultivated at $10\times$ standard PCA concentration (\SI{195}{\micro M}) until chambers were filled, afterwards medium was switched to the desired concentration of PCA for steady state experiments. Microfluidic precultivation in $10\times$ PCA was chosen to ``equilibrate'' cellular metabolism to the carbon source and to reduce experimental time span for filling the microfluidic cultivation chambers. In the dynamic growth experiments (see Fig.~\ref{fig::ov_exp_sim}), cells were directly cultivated under the desired PCA concentration.
\subsection{Live-cell Imaging and Analysis}
\label{matmet::imaging_and_piv}
The microfluidic chip was mounted onto a motorised inverted microscope (Nikon Eclipse Ti, Nikon microscopy, Germany) equipped with an incubator to keep the temperature at $\SI{30}{\celsius}$. Time-lapse phase contrast microscopy images of the growing microcolonies were recorded every $5-10\si{\minute}$ for the growth channel experiments and every $\SI{30}{\minute}$ for the colony spreading experiments over \SI{50}{\hour} of microfluidic cultivation. After the microfluidic cultivation, chambers were manually inspected and selected for analysis. Recordings, in which fabrication inaccuracies led to unstable steady state growth or steady state growth in which differentiation of single cells was not possible anymore, were not further processed. Afterwards, we used the ImageJ PIV plugin implemented by Qingzong Tseng \cite{Tseng31012012} to quantify the velocity field in steady state experiments. In dynamic growth experiments, the total area has been identified using the ``Auto Threshold''-Plugin in ImageJ to identify the outline of the bacterial colony. Hereby, small void spaces inside the colony are added up to the total area as well. Measurements of the total area occupied by the colony have been performed in simulations by subdivision of the simulation domain in a 2D-grid and checking which lattice sides were occupied by at least one cell. We chose quarter the radius of the repulsive interaction between particles as grid constant. We applied a binary closing on the resulting occupation matrix to close small holes inside of the occupation matrix. The area was then defined as the number of occupied lattice sites in the resulting matrix.
\subsection{Fitting Procedure}
\label{ss:fitProc}
To extract values for the model parameters, namely maximum growth rate
$k_{max}=\epsilon u_\infty$ and nutrient diffusion length scale $l_g$,
we fit our model eqns.~\bref{eq::base_model2_ss} to the velocity
profiles $v^{exp}$ measured in experiments. For the uptake function, we tested two different models: Monod-uptake \cite{Monod1949} $u\braces{\hat{g}}=\hat{g}/\braces{1+\hat{g}}$ and Teissier-uptake \cite{teissier1937lois} $u\braces{\hat{g}}=1-\exp\braces{-\hat{g}}$. In one dimension, eqns.~\bref{eq::base_model2_ss} read

\begin{align}
\label{eq:bmss1D_g}
\hat{g}'' &= u\braces{\hat{g}}/l_g^2, \\
\label{eq:bmss1D_v}
v'  &= k_{max} u\braces{\hat{g}},
\end{align}

with the prime denoting spatial deriatives. Insertion of $u\braces{\hat{g}}$ from eq.~\bref{eq:bmss1D_g} into the equation for $v$ \bref{eq:bmss1D_v} leads to 

\begin{equation*}
v' =  k_{max} l_g^2 \hat{g}''.
\end{equation*}

Due to the mirror symmetry around the channel center $x_0\defeq 0$ of our setup, the boundary conditions read $v\braces{x_0} = \hat{g}'\braces{x_0} = 0$.
Integrating once, we obtain 

\begin{equation}
\label{eq:v_dg}
v\braces{x} = k_{max} l_g^2 \hat{g}'\braces{x}.
\end{equation}

We integrate eq.~\bref{eq:bmss1D_g} by using the \verb+odeint()+ method of the python package SciPy.\cite{oliphant2007} The corresponding velocity profile is then given according to eq.~\bref{eq:v_dg}. 
We fit the model solution for the velocity profile to the velocity profiles $v^{exp}$ measured in the experiments by means of a
least-squares optimisation. 
To avoid confusion, we enumerate quantitites belonging to different
experiments with a superscript $i$ in the following, e.g. $v^{exp,i}$
denotes the velocity profile of experiment $i$. We define the cost
function $\Pi$ by the sum of the squared deviations at all points
$x_j^i$ within the fit range measured in experiments: 

\begin{equation}
\Pi = \sum_{ij} \gbraces{[}{ v^{exp,i}\braces{x^i_j} - k_{max} l_g^2 \frac{d\hat{g}^i}{dx}\braces{x^i_j}}{]}^2.
\label{eq:costF}
\end{equation} 

To account for small deviations of the symmetry axis of the experimental flow profiles from the channel center 
the positions $x^i_0$ are also free fit parameters. 
Resulting symmetry axis positions $x^i_0$ deviate only slightly from
the channel center with a relative error  ${x_0}/{2L}<2.5\%$. 
We minimise the cost function $\Pi$ with respect to the parameter $l_g$, the prefactor $\lambda\defeq k_{max} l_g^2$ of $g'$ in eq.~\bref{eq:costF}, and the set of symmetry axis positions $\cbraces{x_0^i}$ and central concentrations $\cbraces{\hat{g}\braces{x_0^i}}$. Thus, for a set of $N$ experiments the fit parameter space is of dimensionality $2+2N$. Minimisation is performed with the \verb+least_squares()+ method of the SciPy python package.\cite{oliphant2007}

As discussed in the main text, the three independent experiments belonging to the PCA-concentration $g_{\infty}=\SI{390}{\micro   M}$ cover almost the complete range of the uptake function as can be seen in Fig.~\ref{fig:u}. To probe the validity of our model, we estimate the model parameters $l_g$ and $k_{max}$ using these experiments, and extrapolate to all other experiments. 
In the remaining experiments only the symmetry points $\cbraces{x_0^i}$ and central concentrations $\cbraces{\hat{g}\braces{x_0^i}}$ are obtained by minimization. 

\begin{figure}[!t]
\begin{center}
\includegraphics[width=\columnwidth]{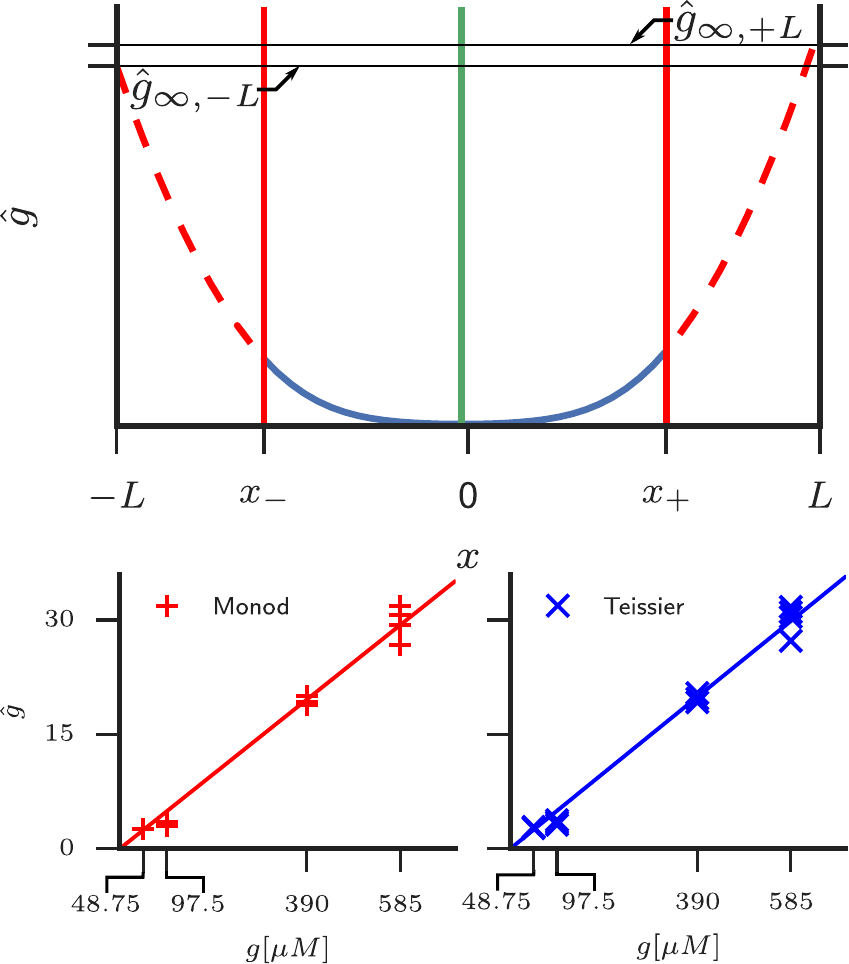}
\caption{{\bf Extrapolation to $\hat{g}_{\infty}$.} {\bf Top:}  Due to the decreasing quality of the PIV data at the channel outlets, the model fit of the velocity profile is constrained to the data between the two points $x_\pm$ (red vertical lines). To get an estimate for $\hat{g}_{\infty}$, we solve the ode  eq.~\bref{eq:bmss1D_g} for the concentration $\hat{g}$ (blue curve) with the fitted model parameters outside the range $x_- \le x \le x_+$ used for the fit (dashed red lines). Due to the slight deviation of the fitted symmetry axis (green vertical) from the center $x=0$, the estimates for $\hat{g}_{\infty,\pm L}$ for $\hat{g}_{\infty}$ differ slightly.\\
{\bf Bottom:} Linear fit $\hat{g} = g/\bar{g}$ of the concentration $\hat{g}$ extrapolated at the channel outlets from our theory and the corresponding concentration in experiment $g$ for Monod and Teissier-uptake.}
\label{fig:gbarEst}
\end{center}
\end{figure}

With the optimal fit solutions for the concentrations $\hat{g}$ at hand, we can estimate the remaining physical parameter, the concentration scale $g_{1/2}$ at which growth and uptake are at half their maximum values. 
To calculate these values in physical units we need to estimate the concentration scale $\bar{g}$ which links between dimensionless theory concentrations $\hat{g}$ and experimental concentrations $g$
via $\hat{g} =g/\bar{g}$. We estimate $\bar{g}$ by comparison
of the feeding concentrations $g_{\infty}$ present at the
channel entries and their model prediction $\hat{g}_{\infty}$. 
We calculate $\hat{g}_{\infty}$ by extrapolation of the
concentration profile $g$ towards the channel entries at $x=\pm L$ (see Fig.~\ref{fig:gbarEst}). 
We expect that this extrapolation only results in coarse estimates for the concentrations $g_{\infty}$ since our model is not strictly valid at the channel entries due to reduced bacterial density. Furthermore, due to the slight deviation of the fitted symmetry axis position $x_0^i$ from the channel center at $x=0$, the extrapolation results in two different concentration values at the entries at $x=\pm L$. We define $\hat{g}_{\infty}$ as their mean.
The extrapolated concentration values $\hat{g}_{\infty}$ show a good agreement with a linear fit $\hat{g}=g/\bar{g}$ as depicted in Fig.~\ref{fig:gbarEst}, bottom. The good match of experimental and theoretical concentrations provides thus an additional consistency check for our model. Table \ref{tab:fitResP} shows a summary of all model parameters resulting from our fitting procedure using Monod or Teissier uptake. 

\begin{table}[tb!]
\small
\caption{Fit results for model parameters obtained from a least-squares minimization}
\label{tab:fitResP}
\begin{center}
\begin{tabular}{cccc}
\hline
uptake&$k_{max} [\si{\hour^{-1}}]$&$l_g [\si{\micro\metre}]$&$g_{1/2} [\si{\micro M}]$\\
\hline
Monod   &$0.26\pm 0.06$&$3.78 \pm 0.17$&$19.9\pm 1.5$\\
Teissier&$0.20 \pm 0.03$&$4.18 \pm 0.14$&$13.6\pm 0.9$\\
\hline
\end{tabular}
\end{center}
Error ranges refer to the square-root of the diagonal entries of the covariance matrix, as reported by \verb+least_squares()+.
\end{table}
\subsection{Simulation Model}
\label{ss:simUP}
\label{matmet::simulation}
\begin{figure}[!t]
\includegraphics[width=\columnwidth]{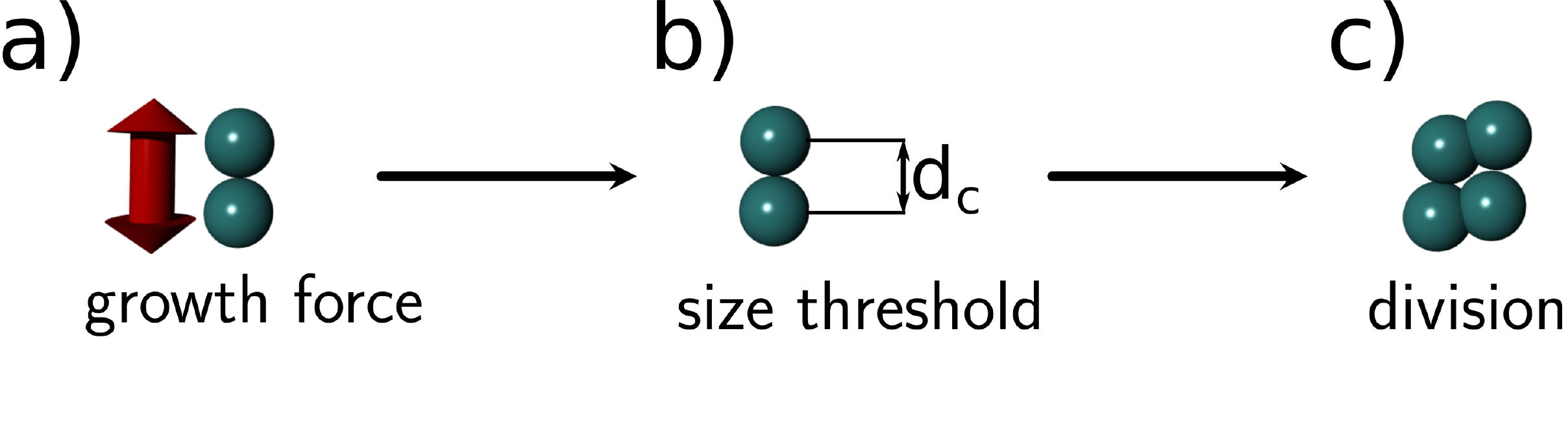}
\caption{\textbf{Schematic of the growth model in simulations.} a) Every cell consists of two point particles which interact with particles of other cells with a purely repulsive interaction, the excluded volume being indicated by spheres. Particles of the same cell repell each other with a growth force $F_{g} =  B/\braces{r+r_0}^2$ until a size threshold $d_c$ is reached. b)-c) At the size threshold a cells divide : two new daughter cells are placed in close vicinity to the mother cell particles. The growth force strength $B$ is proportional to the local nutrient uptake (see eq.~\bref{eq::B_g}) to achieve nutrient-dependent growth.}
\label{fig:SimSchematic}
\end{figure}

We base our simulation on the previously published particle-based model used to study the dynamics of growing systems in various contexts. \cite{basan_dissipative_2011-1,montel_isotropic_2012,epub24238,Podewitz2015}
In this model, cells consist of two particles, that separate due to a
repulsive growth force $F_{g} =  B/\braces{r+r_0}^2$, with force
constants $B$ and $r_0$ (see Fig.~\ref{fig:SimSchematic}). 
Additionally, a friction force between the two cell particles
$\myvec{F}_c=-\gamma_c \myvec{v_c}$ is added, with $\myvec{v_c}$
denoting the relative velocity of the particles constituting one cell.
At a given size threshold $d_c$ the cell divides, and two new daughter
cells are placed in close vicinity to the mother cell. 
The friction constant $\gamma_c$ is chosen large enough to result in
overdamped growth dynamics. Intracell noise is added in a Brownian
Dynamics fashion \cite{frenkel2001understanding}, we denote the
corresponding diffusion constant by $D_c$. The division time of an
isolated cell $\tau_{div}^{sim}$ is then given by 
\begin{equation}
\tau_{div}^{sim} = \frac{\gamma_c}{B} \int_0^{d_c} \braces{r+r_0}^2 dr.  
\label{eq:tdiv}
\end{equation}
Two particles at positions $\myvec{r}_i$ and $\myvec{r}_j$ with $r_{ij}=\vbraces{\myvec{r}_i-\myvec{r}_j}$, belonging
to different cells, interact with each other with a force of magnitude
$F_{cc}=f_0 \braces{1/{r_{ij}^5}-1}-f_1$ directed along the unit
vector $\hat{\myvec{r}}_{ij}=\braces{\myvec{r}_i-\myvec{r}_j}/r_{ij}$. In
this study, we set the attractive component to zero, i.e. $f_1 = 0$
for all simulations. 
Dissipation and fluctuation forces between particles of different
cells are modelled according to the Dissipative Particle
Dynamics-Technique,\cite{frenkel2001understanding} corresponding
friction constant and diffusion constant are denoted by $\gamma_t$ and
$D_t$. A background friction force $\myvec{F}_b=-\gamma_b
\myvec{v}$ with friction constant $\gamma_b$ and $\myvec{v}$, the
velocity in the lab coordinate system, acts on all particles. 
Corresponding background noise is modelled in a Brownian Dynamics fashion \cite{frenkel2001understanding} with diffusion constant $D_b$. All forces are cutoff at a cutoff radius $r_{cut}=\SI{1.1}{\micro\meter}$.
We set the diffusion constants $D_{b,c,t}$ of all particle-particle interactions to $1.21\cdot\si{10^{-3}}\si{\micro\meter^2\hour^{-1}}$.

We integrate the nutrient dynamic eq.~\bref{eq::base_model1} on a square lattice $\braces{x_i,y_j}=h\braces{i,j}$ with lattice constant $h=\SI{1}{\micro\meter}$. We employ a forward-time, central-space finite-difference scheme taking into account spatial varying diffusion constants. \cite{strikwerdaFiniteDiff2004}
In the following, we denote quantities estimated at grid site $\braces{x_i,y_j}$ in simulations with a subscript $ij$, e.g. the concentration $\hat{g}_{ij}= \hat{g}\braces{x_i,y_j}$.
We set the diffusion constant equal to $D_{free}$ at every lattice site with no cell inside and to a smaller value $D_{bulk} \le D_{free}$ at every site with at least one cell present. This models the hindered diffusion of nutrient molecules around and through the cell membrane in a coarse-grained manner. Analytical calculations in \cite{ochoa1994diffusive,wood2002calculation} provide the expression $D_{bulk}=D_{free} \frac{1-\nu}{1+\nu}$ for a two-dimensional array of impermeable cylinders with packing fraction $\nu$. The packing fraction of bacteria of approx. $\nu \approx 0.5$ thus results in $D_{bulk}/D_{free} =1/3$.
The local cell density $\varrho_{ij}$ is obtained from the center of mass coordinates of cells binned on the lattice. 
The cell density $\varrho_{ij}$ is used to calculate the local uptake rate $u_\infty \varrho_{ij} u\!\braces{\hat{g}_{ij}}$ which enters the diffusion eq.~\bref{eq::base_model1}.
We confirmed consistency of results with double or half the grid constant $h$ and agreement with analytical solutions of the one-dimensional equation~\bref{eq:bmss1D_g}.

To transfer the diffusion length scale $l_g$ from our model fit to simulations we choose $u_{\infty} = D_{bulk}/\braces{l_g^2\varrho_{exp.}}$ with $\varrho_{exp.} = \SI{0.66}{\micro\meter^{-2}}$. 
Since the diffusion constant $D_{PCA}$ of our limiting factor PCA is on the order of $\SI{100}{\micro\meter^2\second^{-1}}$ \cite{srinivas2011binary} the concentration profile equilibrates in a timespan $\tau_{diff}=L^2/D$ on the order of seconds. Since bacteria move with a few $\si{\micro\meter\hour^{-1}}$ and divide on a timescale of $\tau_{div} = 1/k_{max}={3-4}\si{\hour}$, the timescale of bacterial dynamics $\tau_{bact}$ is on the order of a few hours. Thus, the concentration profile equilibrates almost instantly on the timescale of bacterial dynamics. Hence, in simulations, it is not necessary to set exactly $D_{free}=D_{PCA}$, any value of $D_{free}$ large enough such that $\tau_{diff}\ll \tau_{bact}$ will result in the same bacterial dynamics (as long as $u_\infty$ is scaled accordingly). For numerical efficiency we chose diffusion constants $D_{bulk} = \SI{2478}{\micro\meter^2/\hour}$ ($D_{bulk} = \SI{1888}{\micro\meter^2/\hour}$ for Teissier-uptake), large enough such that $\tau_{diff} \ll \tau_{div}$.

To reproduce the local growth rate $k_g=k_{max}u\!\braces{\hat{g}}$ of
bacteria (see eq.~\bref{eq::base_model2_ss}) in simulations, we let
the  growth force constant $B$ depend on the local uptake $u\!\braces{\hat{g}}$
(compare eq.~\bref{eq:tdiv})
\begin{equation}
B\braces{\hat{g}} = \frac{\gamma_c  k_{max} u\!\braces{\hat{g}}}{\log\braces{2}} \int_0^{d_c} \braces{r+r_0}^2 dr.
\label{eq::B_g}
\end{equation}
Note, that expression \bref{eq:tdiv} is only valid if the growth force scale $B$ is much larger then the pressure forces the cells are exposed to in the bulk. We can estimate these forces by considering the force balance equation $p'=-\varrho \gamma_b v$ in our one-dimensional model. With vanishing pressure at the channel entries $x=\pm L$ the maximum pressure in the channel center scales with $p\sim \varrho k_{max} \gamma_b L^2$. For the set of simulations as presented in the main text, we aim to stay close to the dynamics as described by model equations \bref{eq::base_model1} and \bref{eq::base_model2} and chose $\gamma_c$ such that $B\braces{g} \gg \varrho k_{max} \gamma_b L^2$, i.e. growth is pressure independent.\footnote{If the scales of local pressure and growth pressure are comparable, growth behaves according to the \textit{homeostatic pressure model}  with a growth rate $k_g \propto p_h-p$, $p_h$ being a species dependent constant called the homeostatic pressure.\cite{basan_homeostatic_2009-1,basan_dissipative_2011-1}}
\begin{figure}[!t]
\begin{center}
\includegraphics[width=\columnwidth]{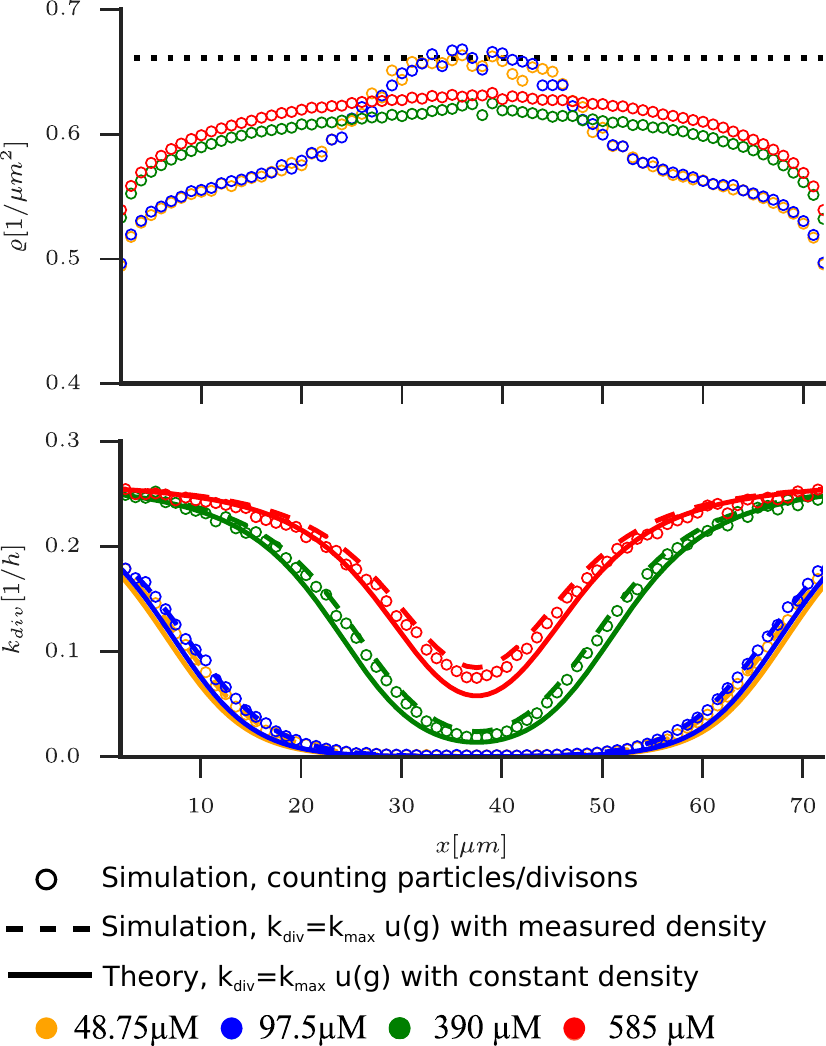}
\end{center}
\caption{{\bf Number density and division rate in simulations.} {\bf Top:} Number density profiles $\varrho$ for steady state simulations at different feeding concentrations (Different colours correspond to different concentrations, see legend at bottom). Horizontal dotted line depicts density value of $\varrho = \SI{0.66}{\micro\meter^{-2}}$ used to calculate the uptake rate $u_{\infty} = D/l_g^2\varrho$ for the simulations from the fitted value for $l_g^2$. {\bf Bottom:} Division rate $k_{div}$, directly measured during simulation by counting division events (empty circles), from $k_{div}=k_{max}\!u\braces{\hat{g}\braces{x}}$ with $\hat{g}(x)$ measured in simulations (dashed lines) and as predicted by theoretical model assuming constant density (continious lines). Depicted results for number density and division rate have been obtained by averaging over time and $y$-direction.}
\label{fig:densKdiv}
\end{figure}
The smooth repulsion potentials result in a non-zero compressibility $K$ scaling with the repulsive force constant $1/K \sim f_0$. Therefore, the number density increases towards the channel center where pressure is largest (see Fig.~\ref{fig:densKdiv} (a)). Due to the non-constant bacterial density $\varrho$ in simulations the concentration profile in simulations is not exactly given by the solution of model eq.~\bref{eq::base_model1} which assumes constant density. Since the concentration profile enters in eq.~\bref{eq::B_g} to determine the local growth rate, model and simulation division rate agree only if the repulsive force constant $f_0$ is large enough to result in only small variations of the number density around $\varrho_{exp}$. For the simulations as presented in the main text we chose a rather large repulsive force constant of $f_0/\gamma_b = \SI{460}{\micro\meter/\hour}$. Thus, separate measurements of the division rate confirm the match between model and simulation, see Fig.~\ref{fig:densKdiv}, bottom.

We also performed growth channel simulations with softer cells and pressure dependent growth which resulted in velocity profiles similar as presented in Fig.~\ref{fig::oneD_flowprof_gconc}(a). However, the match of feeding concentrations in simulations and experiment, as expressed in the linear relationship $\hat{g}=g/\bar{g}$ in Fig.~\ref{fig:gbarEst}(bottom), had a larger error. 
All simulation parameters are summarised in Table \ref{tab:SimP}. 
\section{Acknowledgements}
 Raphael Hornung acknowledges support by the International Helmholtz Research School of Biophysics and Soft Matter (IHRS BioSoft). Microfabrication was performed at Helmholtz Nanoelectronic Facility (HNF) of Forschungszentrum J\"ulich.\cite{albrecht2017hnf} The authors gratefully acknowledge a computing-time grant on the supercomputer JURECA at J\"ulich Supercomputing Centre (JSC\cite{JURECA2016}). We thank Sebastian Schmidt for initiating the cooperation between the institutes IBG-1 and ICS-2.
\section{Funding Statement}
Alexander Gr\"unberger (PD-311) and Dietrich Kohlheyer (VH-NG-1029) are financially supported by the Helmholtz Association.
\section{Author Contributions}
J.E, A.G, R.H, D.K, G.G designed research; R.H, J.E performed data analysis and theoretical modelling ; A.G, C.W 
performed experiment and image analysis; all authors discussed the results and wrote the paper.
\section{Competing Interests}
We have no competing interests.
\section{Data Archiving}
Relevant experimental and simulation data is available in an external repository, \doi{10.5281/zenodo.998797}.

\begin{table*}[!t]
\small
\caption{Summary of simulation parameters}
\label{tab:SimP}
\begin{center}
\begin{tabular}{ccc}
\hline
Parameter & Value & Description \\
\hline
$dt_{DPD}$&$5\times 10^{-5}\si{\hour}$& DPD-integration timestep\\
$h$& \SI{1}{\micro\meter} & Finite-Difference grid constant\\
$dt_{FD}$& $10^{-5}\si{\hour}$ & Finite-Difference timestep\\
$r_0$&\SI{1.1}{\micro\meter}& Growth pressure constant\\
$r_{cut}$&\SI{1.1}{\micro\meter}& Cutoff radius of all pair-potentials\\
$d_c$&\SI{1.1}{\micro\meter}& Size threshold for cell division\\
$r_c$&$1.1\times 10^{-5}\si{\micro\meter}$& \makecell{Distance at which new particles\\ are placed after division}\\
$D_b$&$1.21\times\si{10^{-3}}\si{\micro\meter^2/\hour}$& \makecell{Background noise diffusion constant \\ of bacteria}\\
$D_t$&$1.21\times\si{10^{-3}}\si{\micro\meter^2/\hour}$& \makecell{Intercell noise diffusion constant \\ of bacteria}\\
$D_c$&$1.21\times\si{10^{-3}}\si{\micro\meter^2/\hour}$& \makecell{Intracell noise diffusion constant \\ of bacteria}\\
$\gamma_t/\gamma_b$&$1$& Intercell friction constant\\
$\gamma_c/\gamma_b$&$10^4$& Intracell friction constant\\
$f_0/\gamma_b$&\SI{460}{\micro\meter/\hour}& Repulsive force constant\\
$f_1/\gamma_b$&\SI{0}{\micro\meter/\hour}& Attractive force constant\\
$D_{bulk}$&${2478}(1888)\si{\micro\meter^2/\hour}$& \makecell{Nutrient diffusion constant inside \\ colony for Monod (Teissier)-uptake}\\
\hline
\end{tabular}

Forces are given relative to the background friction constant $\gamma_b$.
\end{center}
\end{table*}

\onecolumn

\bibliographystyle{mybib}
\clearpage
\bibliography{Hornung_RSIF2018}

\begin{thebibliography}{47}
\providecommand{\natexlab}[1]{#1}
\providecommand{\url}[1]{\texttt{#1}}
\providecommand{\urlprefix}{URL }
\expandafter\ifx\csname urlstyle\endcsname\relax
  \providecommand{\doi}[1]{doi:\discretionary{}{}{}#1}\else
  \providecommand{\doi}{doi:\discretionary{}{}{}\begingroup
  \urlstyle{rm}\Url}\fi
\providecommand{\eprint}[2][]{\url{#2}}

\bibitem[{Wang, Robert, Pelletier, Dang, Taddei, Wright et~al.(2010)Wang,
  Robert, Pelletier, Dang, Taddei, Wright, \protect\BIBand{} Jun}]{Wang2010}
Wang P, Robert L, Pelletier J, Dang WL, Taddei F, Wright A, Jun S, 2010 {Robust
  growth of escherichia coli}.
\newblock \emph{Curr. Biol.} 20, 1099--1103.
\newblock \doi{10.1016/j.cub.2010.04.045}

\bibitem[{Taheri-Araghi, Bradde, Sauls, Hill, Levin, Paulsson
  et~al.(2015)Taheri-Araghi, Bradde, Sauls, Hill, Levin, Paulsson, Vergassola,
  \protect\BIBand{} Jun}]{Taheri-Araghi2015}
Taheri-Araghi S, Bradde S, Sauls JT, Hill NS, Levin PA, Paulsson J, Vergassola
  M, Jun S, 2015 {Cell-size control and homeostasis in bacteria}.
\newblock \emph{Curr. Biol.} 25, 385--391.
\newblock \doi{10.1016/j.cub.2014.12.009}

\bibitem[{Gospodarowicz \protect\BIBand{}
  Moran(1976)}]{gospodarowicz1976growth}
Gospodarowicz D, Moran JS, 1976 Growth factors in mammalian cell culture.
\newblock \emph{Annu. Rev. Biochem.} 45, 531--558.
\newblock \doi{10.1146/annurev.bi.45.070176.002531}

\bibitem[{Hirschh{\"a}user, Menne, Dittfeld, West, Mueller-Klieser,
  \protect\BIBand{} Kunz-Schughart(2010)}]{Hirschhaeuser2010}
Hirschh{\"a}user F, Menne H, Dittfeld C, West J, Mueller-Klieser W,
  Kunz-Schughart LA, 2010 {Multicellular tumor spheroids: An underestimated
  tool is catching up again}.
\newblock \emph{J. Biotechnol.} 148, 3--15.
\newblock \doi{10.1016/j.jbiotec.2010.01.012}

\bibitem[{Rumpler, Woesz, Dunlop, van Dongen, \protect\BIBand{}
  Fratzl(2008)}]{Rumpler1173}
Rumpler M, Woesz A, Dunlop JW, van Dongen JT, Fratzl P, 2008 {The effect of
  geometry on three-dimensional tissue growth}.
\newblock \emph{J. R. Soc. Interface} 5, 1173--1180.
\newblock \doi{10.1098/rsif.2008.0064}

\bibitem[{Lovett, Lee, Edwards, \protect\BIBand{} Kaplan(2009)}]{Lovett2009}
Lovett M, Lee K, Edwards A, Kaplan DL, 2009 Vascularization strategies for
  tissue engineering.
\newblock \emph{Tissue Eng. Part B Rev.} 15, 353--370.
\newblock \doi{10.1089/ten.TEB.2009.0085}

\bibitem[{Huh, Hamilton, \protect\BIBand{} Ingber(2011)}]{Huh2011}
Huh D, Hamilton GA, Ingber DE, 2011 {From 3D cell culture to organs-on-chips}.
\newblock \emph{Trends Cell. Biol.} 21, 745--754.
\newblock \doi{10.1016/j.tcb.2011.09.005}

\bibitem[{Bhatia \protect\BIBand{} Ingber(2014)}]{Bhatia2014}
Bhatia SN, Ingber DE, 2014 {Microfluidic organs-on-chips}.
\newblock \emph{Nat. Biotechnol.} 32, 760--772.
\newblock \doi{10.1038/nbt.2989}.
\newblock \eprint{1408.1149}

\bibitem[{Donlan(2002)}]{donlan2002biofilms}
Donlan RM, 2002 {Biofilms: Microbial life on surfaces}.
\newblock \emph{Emerg. Infect. Dis.} 8, 881--890.
\newblock \doi{10.3201/eid0809.020063}

\bibitem[{Stewart(2003)}]{stewart2003diffusion}
Stewart PS, 2003 Diffusion in biofilms.
\newblock \emph{J. Bacteriol.} 185, 1485--1491.
\newblock \doi{10.1128/JB.185.5.1485-1491.2003}

\bibitem[{Wilking, Angelini, Seminara, Brenner, \protect\BIBand{}
  Weitz(2011)}]{Wilking2011}
Wilking JN, Angelini TE, Seminara A, Brenner MP, Weitz DA, 2011 Biofilms as
  complex fluids.
\newblock \emph{MRS Bulletin} 36, 385--391.
\newblock \doi{10.1557/mrs.2011.71}

\bibitem[{Liu, Prindle, Humphries, Gabalda-Sagarra, Asally, Lee
  et~al.(2015)Liu, Prindle, Humphries, Gabalda-Sagarra, Asally, Lee, Ly,
  Garcia-Ojalvo, \protect\BIBand{} S{\"{u}}el}]{Liu2015}
Liu J, Prindle A, Humphries J, Gabalda-Sagarra M, Asally M, Lee DyD, Ly S,
  Garcia-Ojalvo J, S{\"{u}}el GM, 2015 {Metabolic co-dependence gives rise to
  collective oscillations within biofilms}.
\newblock \emph{Nature} 523, 550--554.
\newblock \doi{10.1038/nature14660}.
\newblock \eprint{15334406}

\bibitem[{Whitesides(2006)}]{whitesides2006origins}
Whitesides GM, 2006 The origins and the future of microfluidics.
\newblock \emph{Nature} 442, 368--373.
\newblock \doi{10.1038/nature05058}.
\newblock \eprint{arXiv:1011.1669v3}

\bibitem[{Gr{\"u}nberger, Wiechert, \protect\BIBand{}
  Kohlheyer(2014)}]{grunberger2014single}
Gr{\"u}nberger A, Wiechert W, Kohlheyer D, 2014 Single-cell microfluidics:
  opportunity for bioprocess development.
\newblock \emph{Curr. Opin. Biotechnol.} 29, 15--23.
\newblock \doi{10.1016/j.copbio.2014.02.008}

\bibitem[{Schmid, Kortmann, Dittrich, \protect\BIBand{}
  Blank(2010)}]{schmid2010chemical}
Schmid A, Kortmann H, Dittrich PS, Blank LM, 2010 Chemical and biological
  single cell analysis.
\newblock \emph{Curr. Opin. Biotechnol.} 21, 12--20.
\newblock \doi{10.1016/j.copbio.2010.01.007}

\bibitem[{Kortmann, Blank, \protect\BIBand{} Schmid(2011)}]{kortmann2010single}
Kortmann H, Blank LM, Schmid A, 2011 Single Cell Analytics: An Overview,
  99--122.
\newblock Springer, Berlin.
\newblock \doi{10.1007/10\_2010\_96}

\bibitem[{Mather, Mondrag{\'o}n-Palomino, Danino, Hasty, \protect\BIBand{}
  Tsimring(2010)}]{mather2010streaming}
Mather W, Mondrag{\'o}n-Palomino O, Danino T, Hasty J, Tsimring LS, 2010
  Streaming instability in growing cell populations.
\newblock \emph{Phys. Rev. Lett.} 104, 208101.
\newblock \doi{10.1103/PhysRevLett.104.208101}

\bibitem[{Westerwalbesloh, Gr{\"u}nberger, Stute, Weber, Wiechert, Kohlheyer
  et~al.(2015)Westerwalbesloh, Gr{\"u}nberger, Stute, Weber, Wiechert,
  Kohlheyer, \protect\BIBand{} von Lieres}]{C5LC00646E}
Westerwalbesloh C, Gr{\"u}nberger A, Stute B, Weber S, Wiechert W, Kohlheyer D,
  von Lieres E, 2015 Modeling and cfd simulation of nutrient distribution in
  picoliter bioreactors for bacterial growth studies on single-cell level.
\newblock \emph{Lab Chip} 15, 4177--4186.
\newblock \doi{10.1039/C5LC00646E}

\bibitem[{Cherifi, Jacques, Quessy, \protect\BIBand{}
  Fravalo(2017)}]{Cherifi2017Restriction}
Cherifi T, Jacques M, Quessy S, Fravalo P, 2017 Impact of nutrient restriction
  on the structure of listeria monocytogenes biofilm grown in a microfluidic
  system.
\newblock \emph{Front. Microbiol.} 8, 864.
\newblock \doi{10.3389/fmicb.2017.00864}

\bibitem[{Gr{\"{u}}nberger, Paczia, Probst, Schendzielorz, Eggeling, Noack
  et~al.(2012)Gr{\"{u}}nberger, Paczia, Probst, Schendzielorz, Eggeling, Noack,
  Wiechert, \protect\BIBand{} Kohlheyer}]{grunberger_disposable_2012}
Gr{\"{u}}nberger A, Paczia N, Probst C, Schendzielorz G, Eggeling L, Noack S,
  Wiechert W, Kohlheyer D, 2012 {A disposable picolitre bioreactor for
  cultivation and investigation of industrially relevant bacteria on the single
  cell level}.
\newblock \emph{Lab Chip} 12, 2060--2068.
\newblock \doi{10.1039/c2lc40156h}

\bibitem[{Monod(1949)}]{Monod1949}
Monod J, 1949 The growth of bacterial cultures.
\newblock \emph{Annu. Rev. Microbiol.} 3, 371--394.
\newblock \doi{10.1146/annurev.mi.03.100149.002103}

\bibitem[{Teissier(1937)}]{teissier1937lois}
Teissier G, 1937 Les lois quantitatives de la croissance, volume 455 of
  \emph{Actualit{\'e}s scientifiques et industrielles}, 1--47.
\newblock Hermann {\&} Cie., Paris

\bibitem[{Basan, Prost, Joanny, \protect\BIBand{}
  Elgeti(2011)}]{basan_dissipative_2011-1}
Basan M, Prost J, Joanny JF, Elgeti J, 2011 {Dissipative particle dynamics
  simulations for biological tissues: rheology and competition.}
\newblock \emph{Phys. Biol.} 8, 026014.
\newblock \doi{10.1088/1478-3975/8/2/026014}.
\newblock \eprint{79954495392}

\bibitem[{Farrell, Hallatschek, Marenduzzo, \protect\BIBand{}
  Waclaw(2013)}]{Farrell2013}
Farrell FDC, Hallatschek O, Marenduzzo D, Waclaw B, 2013 Mechanically driven
  growth of quasi-two-dimensional microbial colonies.
\newblock \emph{Phys. Rev. Lett.} 111, 168101.
\newblock \doi{10.1103/PhysRevLett.111.168101}

\bibitem[{Melaugh, Hutchison, Kragh, Irie, Roberts, Bjarnsholt
  et~al.(2016)Melaugh, Hutchison, Kragh, Irie, Roberts, Bjarnsholt, Diggle,
  Gordon, \protect\BIBand{} Allen}]{melaugh2016shaping}
Melaugh G, Hutchison J, Kragh KN, Irie Y, Roberts A, Bjarnsholt T, Diggle SP,
  Gordon VD, Allen RJ, 2016 Shaping the growth behaviour of biofilms initiated
  from bacterial aggregates.
\newblock \emph{PloS one} 11, e0149683.
\newblock \doi{10.1371/journal.pone.0149683}

\bibitem[{Unthan, Gr{\"{u}}nberger, van Ooyen, G{\"{a}}tgens, Heinrich, Paczia
  et~al.(2014)Unthan, Gr{\"{u}}nberger, van Ooyen, G{\"{a}}tgens, Heinrich,
  Paczia, Wiechert, Kohlheyer, \protect\BIBand{} Noack}]{unthan_beyond_2014}
Unthan S, Gr{\"{u}}nberger A, van Ooyen J, G{\"{a}}tgens J, Heinrich J, Paczia
  N, Wiechert W, Kohlheyer D, Noack S, 2014 {Beyond growth rate 0.6: What
  drives Corynebacterium glutamicum to higher growth rates in defined medium}.
\newblock \emph{Biotechnol. Bioeng.} 111, 359--371.
\newblock \doi{10.1002/bit.25103}

\bibitem[{Raffel, Willert, Wereley, \protect\BIBand{}
  Kopenhans(1998)}]{raffel2013particle}
Raffel M, Willert C, Wereley S, Kopenhans J, 1998 {Particle image velocimetry -
  A Practical Guide}, 1--448.
\newblock Springer, Berlin

\bibitem[{Ochoa-Tapia, Stroeve, \protect\BIBand{}
  Whitaker(1994)}]{ochoa1994diffusive}
Ochoa-Tapia JA, Stroeve P, Whitaker S, 1994 Diffusive transport in two-phase
  media: spatially periodic models and maxwell's theory for isotropic and
  anisotropic systems.
\newblock \emph{Chem. Eng. Sci.} 49, 709 -- 726.
\newblock \doi{10.1016/0009-2509(94)85017-8}

\bibitem[{Wood, Quintard, \protect\BIBand{}
  Whitaker(2002)}]{wood2002calculation}
Wood BD, Quintard M, Whitaker S, 2002 Calculation of effective diffusivities
  for biofilms and tissues.
\newblock \emph{Biotechnol. Bioeng.} 77, 495--516.
\newblock \doi{10.1002/bit.10075}

\bibitem[{Allen \protect\BIBand{} Waclaw(2016)}]{allen2016antires}
Allen R, Waclaw B, 2016 Antibiotic resistance: a physicist’s view.
\newblock \emph{Phys. Biol.} 13, 045001.
\newblock \doi{10.1088/1478-3975/13/4/045001}

\bibitem[{Montel, Delarue, Elgeti, Malaquin, Basan, Risler et~al.(2011)Montel,
  Delarue, Elgeti, Malaquin, Basan, Risler, Cabane, Vignjevic, Prost, Cappello,
  \protect\BIBand{} Joanny}]{montel_stress_2011}
Montel F, Delarue M, Elgeti J, Malaquin L, Basan M, Risler T, Cabane B,
  Vignjevic D, Prost J, Cappello G, Joanny JF, 2011 Stress clamp experiments on
  multicellular tumor spheroids.
\newblock \emph{Phys. Rev. Lett.} 107, 188102.
\newblock \doi{10.1103/PhysRevLett.107.188102}.
\newblock \eprint{1111.5814}

\bibitem[{Grant, Wac{\l}aw, Allen, \protect\BIBand{}
  Cicuta(2014)}]{Grant20140400}
Grant MAA, Wac{\l}aw B, Allen RJ, Cicuta P, 2014 The role of mechanical forces
  in the planar-to-bulk transition in growing escherichia coli microcolonies.
\newblock \emph{J. R. Soc. Interface} 11, 20140400.
\newblock \doi{10.1098/rsif.2014.0400}

\bibitem[{Podewitz, Delarue, \protect\BIBand{} Elgeti(2015)}]{Podewitz2015}
Podewitz N, Delarue M, Elgeti J, 2015 Tissue homeostasis: A tensile state.
\newblock \emph{Europhys. Lett.} 109, 58005.
\newblock \doi{10.1209/0295-5075/109/58005}

\bibitem[{Delarue, Hartung, Schreck, Gniewek, Hu, Herminghaus
  et~al.(2016)Delarue, Hartung, Schreck, Gniewek, Hu, Herminghaus,
  \protect\BIBand{} Hallatschek}]{Delarue2016}
Delarue M, Hartung J, Schreck C, Gniewek P, Hu L, Herminghaus S, Hallatschek O,
  2016 Self-driven jamming in growing microbial populations.
\newblock \emph{Nat. Phys.} 12, 762--766.
\newblock \doi{10.1038/nphys3741}

\bibitem[{Farrell, Gralka, Hallatschek, \protect\BIBand{}
  Waclaw(2017)}]{Farrell2017}
Farrell FD, Gralka M, Hallatschek O, Waclaw B, 2017 Mechanical interactions in
  bacterial colonies and the surfing probability of beneficial mutations.
\newblock \emph{J. R. Soc. Interface} 14, 20170073.
\newblock \doi{10.1098/rsif.2017.0073}

\bibitem[{Gr{\"{u}}nberger, Probst, Helfrich, Nanda, Stute, Wiechert
  et~al.(2015)Gr{\"{u}}nberger, Probst, Helfrich, Nanda, Stute, Wiechert, von
  Lieres, N{\"{o}}h, Frunzke, \protect\BIBand{}
  Kohlheyer}]{grunberger2015spatiotemporal}
Gr{\"{u}}nberger A, Probst C, Helfrich S, Nanda A, Stute B, Wiechert W, von
  Lieres E, N{\"{o}}h K, Frunzke J, Kohlheyer D, 2015 {Spatiotemporal microbial
  single-cell analysis using a high-throughput microfluidics cultivation
  platform}.
\newblock \emph{Cytometry A} 87, 1101--1115.
\newblock \doi{10.1002/cyto.a.22779}

\bibitem[{Gr{\"{u}}nberger, Probst, Heyer, Wiechert, Frunzke, \protect\BIBand{}
  Kohlheyer(2013)}]{gruenberger2013microfluidic}
Gr{\"{u}}nberger A, Probst C, Heyer A, Wiechert W, Frunzke J, Kohlheyer D, 2013
  Microfluidic picoliter bioreactor for microbial single-cell analysis:
  Fabrication, system setup, and operation.
\newblock \emph{J. Vis. Exp.} 1 -- 11.
\newblock \doi{10.3791/50560}

\bibitem[{Tseng, Duchemin-Pelletier, Deshiere, Balland, Guillou, Filhol
  et~al.(2012)Tseng, Duchemin-Pelletier, Deshiere, Balland, Guillou, Filhol,
  \protect\BIBand{} Théry}]{Tseng31012012}
Tseng Q, Duchemin-Pelletier E, Deshiere A, Balland M, Guillou H, Filhol O,
  Théry M, 2012 Spatial organization of the extracellular matrix regulates
  cell–cell junction positioning.
\newblock \emph{Proc. Natl. Acad. Sci. USA} 109, 1506--1511.
\newblock \doi{10.1073/pnas.1106377109}

\bibitem[{Oliphant(2007)}]{oliphant2007}
Oliphant TE, 2007 Python for scientific computing.
\newblock \emph{Comput. Sci. Eng.} 9, 10--20.
\newblock \doi{10.1109/MCSE.2007.58}

\bibitem[{Montel, Delarue, Elgeti, Vignjevic, Cappello, \protect\BIBand{}
  Prost(2012)}]{montel_isotropic_2012}
Montel F, Delarue M, Elgeti J, Vignjevic D, Cappello G, Prost J, 2012 Isotropic
  stress reduces cell proliferation in tumor spheroids.
\newblock \emph{New J. Phys.} 14, 055008.
\newblock \doi{10.1088/1367-2630/16/11/115005}

\bibitem[{Marel, Podewitz, Zorn, Rädler, \protect\BIBand{}
  Elgeti(2014)}]{epub24238}
Marel AK, Podewitz N, Zorn M, Rädler JO, Elgeti J, 2014 Alignment of cell
  division axes in directed epithelial cell migration.
\newblock \emph{New J. Phys.} 16, 115005.
\newblock \doi{10.1088/1367-2630/16/11/115005}

\bibitem[{Frenkel \protect\BIBand{} Smit(2001)}]{frenkel2001understanding}
Frenkel D, Smit B, 2001 {Understanding Molecular Simulation: From Algorithms to
  Applications}.
\newblock Academic press, London, 2nd edition

\bibitem[{Strikwerda(2004)}]{strikwerdaFiniteDiff2004}
Strikwerda J, 2004 Finite Difference Schemes and Partial Differential
  Equations, p. 163, eq. (6.5.1).
\newblock Society for Industrial and Applied Mathematics, 2nd edition.
\newblock \doi{10.1137/1.9780898717938}

\bibitem[{Srinivas, King, Howard, \protect\BIBand{}
  Monrad(2011)}]{srinivas2011binary}
Srinivas K, King JW, Howard LR, Monrad JK, 2011 {Binary diffusion coefficients
  of phenolic compounds in subcritical water using a chromatographic peak
  broadening technique}.
\newblock \emph{Fluid Phase Equilib.} 301, 234--243.
\newblock \doi{10.1016/j.fluid.2010.12.003}

\bibitem[{Basan, Risler, Joanny, {Sastre Garau}, \protect\BIBand{}
  Prost(2009)}]{basan_homeostatic_2009-1}
Basan M, Risler T, Joanny JF, {Sastre Garau} X, Prost J, 2009 {Homeostatic
  competition drives tumor growth and metastasis nucleation}.
\newblock \emph{HFSP J.} 3, 265--272.
\newblock \doi{10.2976/1.3086732}.
\newblock \eprint{0902.1730}

\bibitem[{{Forschungszentrum Jülich GmbH.}(2017)}]{albrecht2017hnf}
{Forschungszentrum Jülich GmbH}, 2017 {HNF-Helmholtz Nano Facility}.
\newblock \emph{Journal of large-scale research facilities JLSRF} 3, {A112}.
\newblock \doi{10.17815/jlsrf-3-158}

\bibitem[{{J\"ulich Supercomputing Centre.}(2016)}]{JURECA2016}
{J\"ulich Supercomputing Centre}, 2016 {JURECA: General-purpose supercomputer
  at J\"ulich Supercomputing Centre.}
\newblock \emph{Journal of large-scale research facilities JLSRF} 2, {A62}.
\newblock \doi{10.17815/jlsrf-2-121}

\end{thebibliography}

\end{document}